\documentclass{iopart}
\usepackage{iopams}
\usepackage{graphicx}

\renewcommand{\Re}{\mathop{\mathrm{Re}}\nolimits}

\begin{document}

\title[Electron spin resonance in bulk topological 
insulators]
{Theory of electron spin resonance in bulk topological 
insulators Bi$_2$Se$_3$, Bi$_2$Te$_3$ and Sb$_2$Te$_3$}

\author{O Ly$^1$ and D M Basko$^2$}
\address{$^1$ Institut de Physique et Chimie des Mat\'eriaux de Strasbourg, 67000 Strasbourg, France}
\address{$^2$ Laboratoire de Physique et Mod\'elisation des Milieux Condens\'es, Universit\'e Grenoble Alpes and CNRS, B.P. 166, 38042 Grenoble, France}
\ead{ousmane.ly@ipcms.cnrs.fr}

\begin{abstract}
We report a theoretical study of electron spin resonance in bulk topological insulators, such as Bi$_2$Se$_3$, Bi$_2$Te$_3$ and Sb$_2$Te$_3$. Using the effective four-band model, we find the electron energy spectrum in a static magnetic field and determine the response to electric and magnetic dipole perturbations, represented by oscillating electric and magnetic fields perpendicular to the static field. We determine the associated selection rules and calculate the absorption spectra. This enables us to separate the effective orbital and spin degrees of freedom and to determine the effective $g$~factors for electrons and holes.
\end{abstract}

\submitto{\JPCM}

%\maketitle

\section{Introduction}
\paragraph{}

The discovery of Bi$_2$Se$_3$, Bi$_2$Te$_3$ and Sb$_2$Te$_3$ being
topological insulators~\cite{Zhang2009,Xia2009,Chen2009,Hsieh2009}
has greatly stimulated research on these materials, the main object
of interest being the existence of topologically protected conducting
surface states~\cite{Bernevig2013}. The interest in bulk properties
of these materials is driven by their high performance as 
thermoelectrics~\cite{Hor2009,Kadel2011,Osterhage2014}. A powerful
tool to probe electronic properties of solids is magneto-optical
spectroscopy which was recently applied to bulk Bi$_2$Se$_3$, where
optical transitions between electronic Landau levels were
observed~\cite{Orlita2015}.

The standard textbook picture of non-relativistic electron motion
in a static uniform magnetic field involves Landau quantization of 
the  orbital motion in the plane perpendicular to the field, and 
Zeeman   splitting of the Landau levels according to the spin 
projection.
The strength of the Zeeman splitting is characterized by the 
effective $g$~factor. The orbital and spin degrees of freedom can be
addressed separately by applying oscillating electric or magnetic
field perpendicular to the static field, resulting respectively in 
transitions between orbital Landau levels (cyclotron resonance) or 
Zeeman sublevels (electron spin resonance, ESR, also called electron 
paramagnetic resonance).
The ESR spectroscopy \cite{Zavoisky1945,Abragam1970} is a powerful
tool for studying impurity electron spins, as well as those of
conduction electrons
\cite{Griswold1952,Dyson1955,Lampe1966,Seck1997,Wilamowski2002,Rieger2007}.
An ESR experiment in Bi$_2$Se$_3$ has recently been
reported~\cite{Wolos2013}.

The simple picture of separation between orbital and spin degrees of
freedom breaks down if a strong spin-orbit coupling is
present. Indeed, solution of the effective Schr\"odinger equation
in the static magnetic field~$B_z$ gives a set of $B_z$-dependent
energy levels~\cite{Liu2010}, and it is not obvious how to separate
the quantum numbers into effective orbital and spin ones, and how to
define the effective $g$~factors. At the same time, the response to
physical perturbations, such as oscillating electric and magnetic 
fields, corresponding to the cyclotron resonance and ESR experiments,
can be determined unambiguously. Then, one can try to
analyze the corresponding transitions, with the aim of determining
the effective orbital and spin degrees of freedom, and extracting
the effective $g$~factors. This is the subject of the present work.

\section{The model}
\setlength{\mathindent}{3cm}
%\section{Model Hamiltonian }
%\paragraph{}
We use the effective model for 3D topological insulators Bi$_2$Se$_3$, Bi$_2$Te$_3$ and Sb$_2$Te$_3$ proposed in~\cite{Zhang2009}. The full microscopic derivation of this model is given in \cite{Liu2010}. Neglecting terms of degree $k^3$ and higher, we start from the following Hamiltonian in the absence of the magnetic field:
\begin{equation}\label{Hk=}
H_\mathbf{k}=\varepsilon(\mathbf{k})+\pmatrix{%
 \mathcal{M}(\mathbf{k}) & \mathcal{B}_0k_z & 0 & \mathcal{A}_0k_- \cr
 \mathcal{B}_0k_z & -\mathcal{M}(\mathbf{k}) & \mathcal{A}_0k_- & 0 \cr
 0 & \mathcal{A}_0k_+ & \mathcal{M}(\mathbf{k}) & -\mathcal{B}_0k_z \cr
 \mathcal{A}_0k_+ & 0 & -\mathcal{B}_0k_z & -\mathcal{M}(\mathbf{k})\cr},
\end{equation}
leading to the energy spectrum :  
\begin{equation}\label{Ek=}
	E_\mathbf{k}=\varepsilon(\mathbf{k})\pm \sqrt{\mathcal{M}(\mathbf{k})^2+	\mathcal{A}_0^2(k_x^2+k_y^2)+\mathcal{B}_0^2k_z^2} ,
\end{equation}
where
$\varepsilon(\mathbf{k})=C_1k_z^2+C_2(k_x^2+k_y^2)$ (we omit the
constant term~$C_0$),
$\mathcal{M}(\mathbf{k})=M_0+M_1k_z^2+M_2(k_x^2+k_y^2)$,
and we denoted $k_{\pm}=k_x\pm i k_y$.
The band structure parameters $C_{1,2},M_{0,1,2},\mathcal{A}_0,\mathcal{B}_0$ have been estimated in~\cite{Liu2010} from a combination of \textit{ab initio} calculations and $\mathbf{k}\cdot\mathbf{p}$ perturbation theory. Somewhat different values of parameters $C_2,M_0,M_2,\mathcal{A}_0$ have been proposed in~\cite{Orlita2015} to match magneto-optical spectroscopy data for Bi$_2$Se$_3$.
Hamiltonian~(\ref{Hk=}) is invariant under the time reversal: %operation:
\begin{equation}\label{UT=}
U_\mathrm{T}H_{-\mathbf{k}}^*U_\mathrm{T}^{-1}=H_\mathbf{k},\quad
U_\mathrm{T}=\pmatrix{0 & 0 & 1 & 0 \cr 0 & 0 & 0 & 1 \cr
-1 & 0 & 0 & 0 \cr 0 & -1 & 0 & 0}.
\end{equation}

The Hamiltonian in the presence of an external magnetic field is
obtained by the Peierls substitution
$\mathbf{k}\rightarrow\bpi=
-i\bnabla+(e/\hbar{c})\mathbf{A}$
in (\ref{Hk=}), where we assume the electron charge to be $-e$,
and by including the gauge-invariant Zeeman-type terms~\cite{Liu2010}:
\begin{equation}\label{HZ=}
H_\mathrm{Z}=\frac{\mu_\mathrm{B}}2\pmatrix{
g_{1z}B_z & 0 & g_{1p}B_- & 0 \cr 0 & g_{2z}B_z & 0 & g_{2p}B_- \cr
g_{1p}B_+ & 0 & -g_{1z}B_z & 0 \cr 0 & g_{2p}B_+ & 0 & -g_{2z}B_z},
\end{equation}
where $\mu_\mathrm{B}$ is the Bohr magneton,
%$\mu_\mathrm{B}=e\hbar/(2m_0c)$ is the Bohr magneton ($m_0$~is the
%free electron mass),
$B_\pm=B_x\pm{i}B_y$, and the $g$-factors 
$g_{1z},g_{2z},g_{1p},g_{2p}$ have been estimated in~\cite{Liu2010}.
For a constant homogeneous field $B_z$ along the $z$ direction,
we use the Landau gauge $A_x=-B_zy$, $A_y=A_z=0$.

The electric dipole perturbation is obtained by applying a uniform
oscillating electric field in the direction perpendicular to the
constant magnetic field. As the system is isotropic in the $xy$
plane, we can choose the $x$ direction without the loss of
generality. This corresponds to the following total vector potential:
\begin{equation}\label{eldip=}
A_x=-B_zy+\frac{c\mathcal{E}_x}{\rmi\omega}\,\rme^{-\rmi\omega{t}}
+\mathrm{c.c.},\quad A_y=A_z=0,
\end{equation}
where $\mathcal{E}_x$ is the electric field amplitude, and ``c.c.''
stands for complex conjugate. The magnetic dipole perturbation
(without electric quadrupole contribution) corresponds to the vector
potential~\cite{Jackson}
\begin{equation}\label{magdip=}
A_x=-B_zy,\quad
A_y=-\frac{B_xz}{2}\,\rme^{-\rmi\omega{t}}+\mathrm{c.c.},\quad
A_z=\frac{B_xy}{2}\,\rme^{-\rmi\omega{t}}+\mathrm{c.c.},
\end{equation}
which gives a uniform oscillating magnetic field along~$x$, as well
as a position-dependent oscillating electric field, 
$-(1/c)\partial\mathbf{A}/\partial{t}$. Indeed, by Faraday's law,
a time-dependent magnetic field is necessarily accompanied by an
electric field.

In the following, we will calculate the energy absorbed by the
system due to each of the two perturbations. We will use the
general expression of the linear response theory for an
electronic system whose single-particle Hamiltonian $H$ contains
a stationary part~$H_0$ and a monochromatic perturbation,
\begin{equation}
H(t)={H}_0+{F}\rme^{-\rmi\omega{t}}+{F}^\dag\rme^{\rmi\omega{t}}.
\end{equation}
The power, absorbed by the system, is given by the sum over
single-electron transitions $\beta\to\alpha$ whose rates are
obtained from the Fermi Golden Rule~\cite{LL3}:
\begin{equation}\label{FGR=}
P(\omega)=2\pi\omega\sum_{\alpha,\beta}|F_{\alpha\beta}|^2
(f_\beta-f_\alpha)\,\delta(E_\alpha-E_\beta-\hbar\omega),
\end{equation}
where $\alpha,\beta$ label the eigenstates of ${H}_0$ with
energies $E_\alpha,E_\beta$ and occupations $f_\alpha,f_\beta$,
and $F_{\alpha\beta}$~is the matrix element of the perturbation
between these states.

\section{Results}

\subsection{Landau levels}

Landau levels for the described model were found in~\cite{Liu2010}.
It is convenient to define the raising and lowering operators
$\hat{a}^\dag=\pi_+l_B/\sqrt{2}$, $\hat{a}=\pi_-l_B/\sqrt{2}$, 
where $l_B=\sqrt{\hbar{c}/(eB_z)}$ is the magnetic length.
%Then the Hamiltonian takes the form
%\begin{eqnarray}
%\hat{H}_0&=&\hat\varepsilon+\pmatrix{%
%\hat{\mathcal{M}} & \mathcal{B}_0(-i\partial_z) & 0 & \Omega_B\hat{a} \cr
%\mathcal{B}_0(-i\partial_z) & -\hat{\mathcal{M}} & \Omega_B\hat{a} & 0 \cr
%0 & \Omega_B\hat{a}^{\dag} & \hat{\mathcal{M}} & \mathcal{B}_0\,i\partial_z \cr
%\Omega_B\hat{a}^{\dag} & 0 & \mathcal{B}_0\,i\partial_z & -\hat{\mathcal{M}}}
%+\nonumber\\
%&&{}+\frac{\mu_\mathrm{B}B_z}{2}\pmatrix{g_{1z} & 0 & 0 & 0\cr 0 & g_{2z} & 0 & 0\cr
%0 & 0 & -g_{1z} & 0 \cr 0 & 0 & 0 & -g_{2z}},
%\label{H0=}
%\end{eqnarray}
%where we denoted $\Omega_B\equiv\sqrt{2}\mathcal{A}_0/l_B$,
%\begin{eqnarray}
%\hat\varepsilon&=&\frac{2C_2}{l_B^2}
%\left(\hat{a}^{\dag}\hat{a}+\frac{1}{2}\right)-C_1\partial_z^2,\\
%\hat{\mathcal{M}}&=&M_0+\frac{2M_2}{l_B^2}
%\left(\hat{a}^{\dag}\hat{a}+\frac{1}{2}\right)-M_1\partial_z^2.
%\end{eqnarray}
The wave functions can be sought in the form
\begin{equation}\label{wf4=}
\psi(x,y,z)=\frac{\rme^{\rmi{k}_xx+\rmi{k}_zz}}{\sqrt{L_xL_z}}
\pmatrix{ b_{n1}\,\phi_{n-1}\cr b_{n2}\,\phi_{n-1}\cr
b_{n3}\,\phi_n\cr b_{n4}\,\phi_n\cr},
\end{equation}
where $L_{x,y,z}$ is the size of the sample in the corresponding
direction, $n\geqslant{0}$ is the Landau level index,
$\phi_n=\phi_n(k_xl_B^2-y)$ are the normalized harmonic oscillator
wave functions, and  $b_{nj}$, $j=1,\ldots,4$~are some unknown
coefficients ($b_{01}$ and $b_{02}$ must vanish as $\phi_{-1}$
is not defined).
In this representation, $\hat{a}\phi_n=\sqrt{n}\,\phi_{n-1}$,
%and $\hat{a}^\dagger\phi_n=\sqrt{n+1}\,\phi_{n+1}$,
so the coefficients $b_{nj}$ satisfy the $4\times{4}$ eigenvalue
problem $(H_n-E)b_n=0$ with the matrix $H_n$ given by
\begin{eqnarray}
H_n&=&\pmatrix{%
\varepsilon_n+\mathcal{M}_n & \mathcal{B}_0k_z & 0 &
\sqrt{2n}\mathcal{A}_0/l_B \cr
\mathcal{B}_0k_z & \varepsilon_n-\mathcal{M}_n &
\sqrt{2n}\mathcal{A}_0/l_B & 0 \cr
0 & \sqrt{2n}\mathcal{A}_0/l_B & \varepsilon_n+\mathcal{M}_n &
 -\mathcal{B}_0k_z \cr
\sqrt{2n}\mathcal{A}_0/l_B & 0 & -\mathcal{B}_0k_z &
\varepsilon_n-\mathcal{M}_n
}+\nonumber\\
&&+\frac{\mu_\mathrm{B}B_z}{2}\pmatrix{
\tilde{g}_{1z} & 0 & 0 & 0\cr 0 & \tilde{g}_{2z} & 0 & 0\cr
0 & 0 & -\tilde{g}_{1z} & 0 \cr 0 & 0 & 0 & -\tilde{g}_{2z}},
\label{Hn=}
\end{eqnarray}
where we denoted 
$\varepsilon_n=2nC_2/l_B^2+C_1k_z^2$,
$\mathcal{M}_n=M_0+2nM_2/l_B^2+M_1k_z^2$,
$\tilde{g}_{1z}=g_{1z}-4m_0(C_2+M_2)/\hbar^2$,
$\tilde{g}_{2z}=g_{2z}-4m_0(C_2-M_2)/\hbar^2$, and $m_0$ stands for
the free electron mass entering the Bohr magneton
$\mu_\mathrm{B}=e\hbar/(2m_0c)$.

The matrix $H_n$ from (\ref{Hn=}) cannot be diagonalized 
analytically, however, some general properties of the spectrum can
be established:
\begin{enumerate}
\item
the energy does not depend on~$k_x$, which determines the usual
degeneracy $L_xL_y/(2\pi{l}_B^2)$ of each Landau level;
\item
for each $k_z$, there are four energy levels,
$E_{nj}(k_z)$, $j=1,\ldots,4$, for $n>0$,
while for $n=0$ there are only two levels;
\item
at $k_z=0$, the Hamiltonian splits in two decoupled $2\times{2}$
blocks.
\end{enumerate}
To determine the dispersion for $k_z\neq{0}$, one can use
perturbation theory in
$v_z=(1/\hbar)(\partial{H}_\mathbf{k}/\partial{k}_z)$, which has
no diagonal matrix elements within the blocks at $k_z=0$. Thus,
the dispersion has no linear term in $k_z$, so one can approximate
it by a quadratic one near $k_z=0$:
$E_{nj}(k_z)\approx{E}_{nj}(0)+\alpha_{nj}k_z^2$.
As a result, the joint density of states has a series of
square-root-type singularities near frequencies corresponding
to transition energies $E_{n'j'}(0)-E_{nj}(0)$ at $k_z=0$:
\begin{eqnarray}
P(\omega)&=&2\pi\hbar\omega\sum_{n,j,n',j'}\frac{L_xL_y}{2\pi{l}_B^2}
\int\frac{L_z\,\rmd{k}_z}{2\pi}
\left|F_{nj,n'j'}(k_z)\right|^2\times\nonumber\\
&&{}\times\delta\left(E_{n'j'}(k_z)-E_{nj}(k_z)-\hbar\omega\right)
\approx\nonumber\\
&\approx&\sum_{n,j,n',j'}
\frac{\omega\left|F_{n,j;n',j'}(0)\right|^2L_xL_yL_z/(2\pi{l}_B^2)}%
{\sqrt{|\alpha_{n'j'}-\alpha_{nj}|(\hbar\omega+E_{nj}(0)-E_{n'j'}(0))}},
\label{Psqrt=}
\end{eqnarray}
where we took into account the fact that $k_z$ is not changed under
the excitation, and the summation is over levels $n,j$ which are
filled, and $n',j'$ which are empty. In a real sample, these
singularities are smeared due to level broadening (e.~g., by
disorder), so experimental spectra consist of broadened peaks
located near transition energies $E_{n'j'}(0)-E_{nj}(0)$
(asymmetric peak
smearing may also lead to a shift of the maximum frequency).
Their intensities are determined by the perturbation matrix elements
between states at $k_z=0$.\footnote{
One should be cautious when applying this argument to intraband
transitions at low fields. Indeed, at $B_z\to{0}$, all
$\alpha_{n,j}\to{C}_1\pm[M_1+\mathcal{B}_0^2/(2M_0)]$, so the
denominator in \eref{Psqrt=} vanishes. The peak shape is then
determined mostly by the broadening and by higher-order in~$k_z$
terms in $E_{nj}(k_z)$.}

At $k_z=0$, Hamiltonian~(\ref{Hn=}) consists of two independent
$2\times{2}$ blocks. We label the outer/inner block by
$\sigma=\uparrow,\downarrow=\pm{1}$, respectively.
Besides $n$ and $\sigma$, the levels with $n>0$ are also
labeled by the band index $\lambda=\pm{1}$. Their energies are
given by
\begin{eqnarray}\nonumber
&&E_{n>0,\sigma,\lambda}=\lambda\sqrt{\frac{2n}{l_B^2}\mathcal{A}_0^2
+\left[M_0+\frac{2nM_2}{l_B^2} -
\frac{\sigma}{l_B^2}\left(C_2-\frac{g_{1z}+g_{2z}}{8m_0/\hbar^2}\right)\right]^2}\\ 
&&\hspace*{2cm}{}+\frac{2nC_2}{l_B^2} -
\frac{\sigma}{l_B^2} \left(M_2-\frac{g_{1z}-g_{2z}}{8m_0/\hbar^2}\right),
\label{LLn=}\\
&&E_{n=0,\sigma=\pm{1}}=\frac{C_2}{l_B^2}\mp\left(M_0+\frac{M_2}{l_B^2}\right)
-g_{2z,1z}\,\frac{\mu_\mathrm{B}B_z}{2}.\label{LL0=}
\end{eqnarray}
%The wave functions at $k_z=0$ can be written as
%\begin{equation}
%b_{n,\uparrow,\pm{1}}=
%\pmatrix{ \xi_{n\uparrow} \cr 0 \cr 0 \cr \eta_{n\uparrow}},
%\pmatrix{ \eta_{n\uparrow} \cr 0 \cr 0 \cr -\xi_{n\uparrow}};\quad
%b_{n,\downarrow,\pm{1}}=
%\pmatrix{ 0\cr \eta_{n\downarrow} \cr \xi_{n\downarrow} \cr 0},
%\pmatrix{ 0\cr -\xi_{n\downarrow} \cr \eta_{n\downarrow} \cr 0}.
%\end{equation}
%If one neglects $\tilde{g}_{1z},\tilde{g}_{2z}$, then
%$\xi_{n\downarrow}=\xi_{n\uparrow}$, $\eta_{n\downarrow}=\eta_{n\uparrow}$.
%For $M_0<0$, we have $\eta_{0\sigma}=1$, $\xi_{0\sigma}=0$.
Corrections to the Landau level energies,
$\alpha_{n\sigma\lambda}k_z^2$, can then be obtained by perturbation
theory for Hamiltonian~(\ref{Hn=}) in $C_1k_z^2$, $M_1k_z^2$ (to the
first order), and in $\mathcal{B}_0k_z$ (to the second order).
%This determines the coefficients $\alpha_{n,\sigma,\lambda}$
%\textbf{(check)}:
%\begin{eqnarray}
%\alpha_{n,\sigma,\lambda}=&C_1
%+\lambda{M}_1(\xi_{n\sigma}^2-\eta_{n\sigma}^2)+\nonumber\\
%&+\mathcal{B}_0^2\,\frac{(\xi_{n\uparrow}\xi_{n\downarrow}
%+\eta_{n\uparrow}\eta_{n\downarrow})^2}{E_{n,\sigma,\lambda}
%-E_{n,-\sigma,-\lambda}}+\mathcal{B}_0^2\,
%\frac{(\xi_{n\downarrow}\eta_{n\uparrow}
%-\eta_{n\downarrow}\xi_{n\uparrow})^2}{E_{n,\sigma,\lambda}
%-E_{n,-\sigma,\lambda}}.
%\end{eqnarray}

We emphasize that at this point, $\sigma$~is just a quantum number
introduced formally in order to distinguish between the two blocks
at $k_z=0$. Its relation to the spin degree of freedom will be 
established below by studying the response to physical perturbations. 
Note, however, that the two sectors are exchanged by the time 
reversal operation~(\ref{UT=}), as is also the case for the real 
spin.

\subsection{Electric dipole perturbation}

The perturbation, corresponding to the vector potential~(\ref{eldip=}),
is given by
\begin{equation}\label{Fe=}
F_\mathrm{e}=\frac{\rmi e\mathcal{E}_x}\omega\,v_x,\quad
\mathbf{v}\equiv\frac{1}\hbar\left.
\frac{\partial{H}_\mathbf{k}}{\partial\mathbf{k}}
\right|_{\mathbf{k}\to\bpi},
\end{equation}
and the absorbed power found from~\eref{FGR=} is related to the
real part of the optical conductivity $\sigma_{ij}(\omega)$ as
\begin{equation}
P(\omega)=2\Re\sigma_{xx}(\omega)|\mathcal{E}_x|^2L_xL_yL_z.
\end{equation}
At $k_z=0$, the evaluation of the matrix elements of $F_\mathrm{e}$
is straightforward, but the resulting
expressions are cumbersome, so we do not give them here.
%It is important that the perturbation 
$F_\mathrm{e}$~has no matrix elements between
different $\sigma$ sectors, so the selection rules are
\begin{equation}\label{electricSR=}
%F^e_{n\sigma\lambda,n'\sigma'\lambda'}\propto
%\delta_{\sigma'\sigma}\delta_{n',n+1}
%+\delta_{\sigma'\sigma}\delta_{n',n-1}.
n,\sigma,\lambda\to n\pm{1},\sigma,\lambda',
\end{equation}
with no restriction on $\lambda,\lambda'$ (note, however, that
for interband transitions, some matrix elements can be much
stronger than others, as discussed in the end of
section~\ref{sec:Discussion}).
Representing the linear polarization as a sum of two circular
polarizations, one can see that the plus/minus sign corresponds
to the counterclockwise/clockwise rotating electric field
(left/right circular polarization), respectively. 

\begin{figure}
\includegraphics[width=10cm]{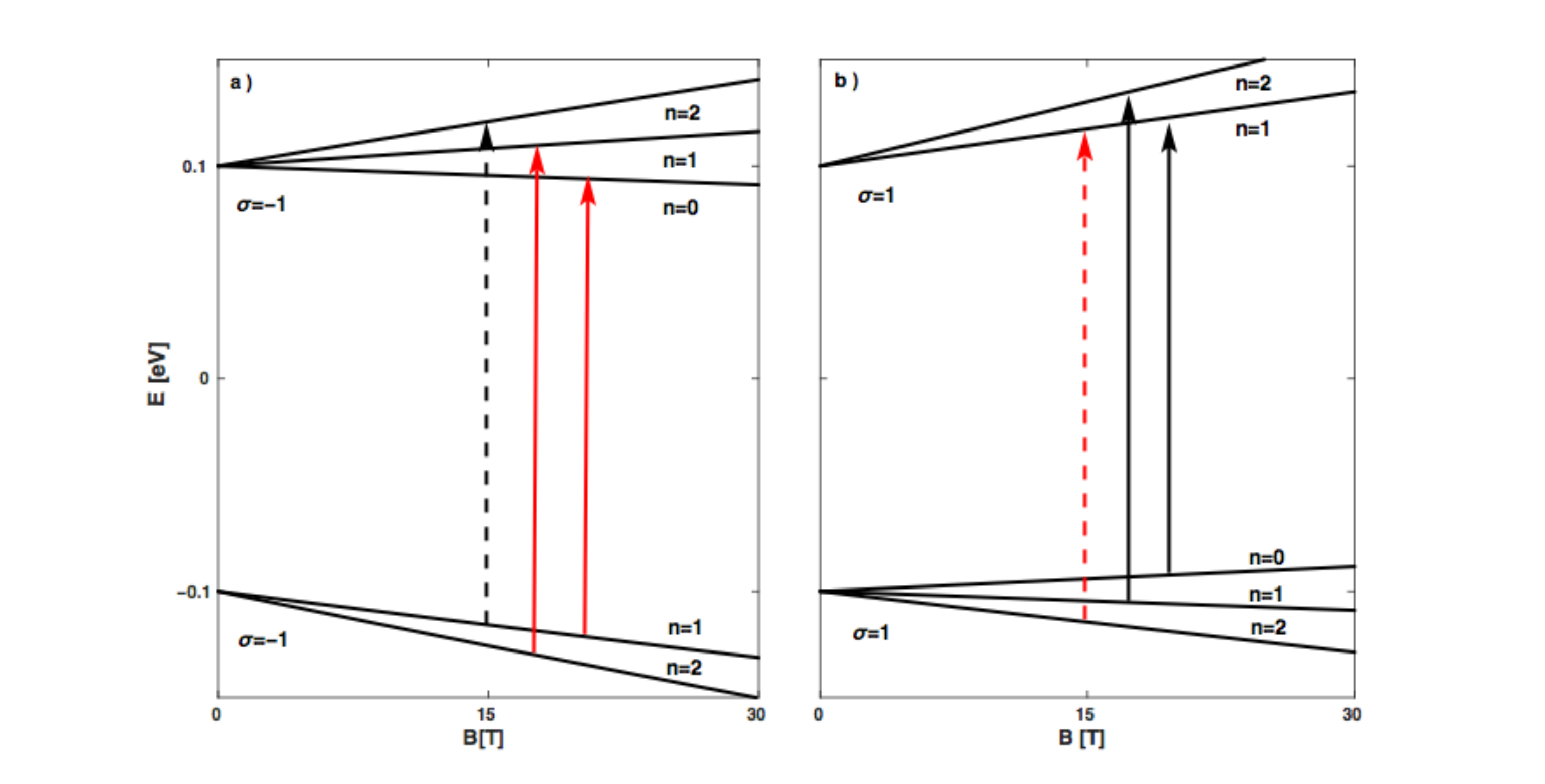}
\caption{\label{fig:TransitionsE}
Interband transverse electric dipole transitions for the first three
Landau levels for $\sigma=\pm{1}$ [panels (a)
and (b), respectively].
Solid and dashed arrows show dominant and subdominant transitions.
Left and right circular polarizations are shown by black and red arrows, respectively.
}
\end{figure}

\begin{figure}
\includegraphics[width=6cm]{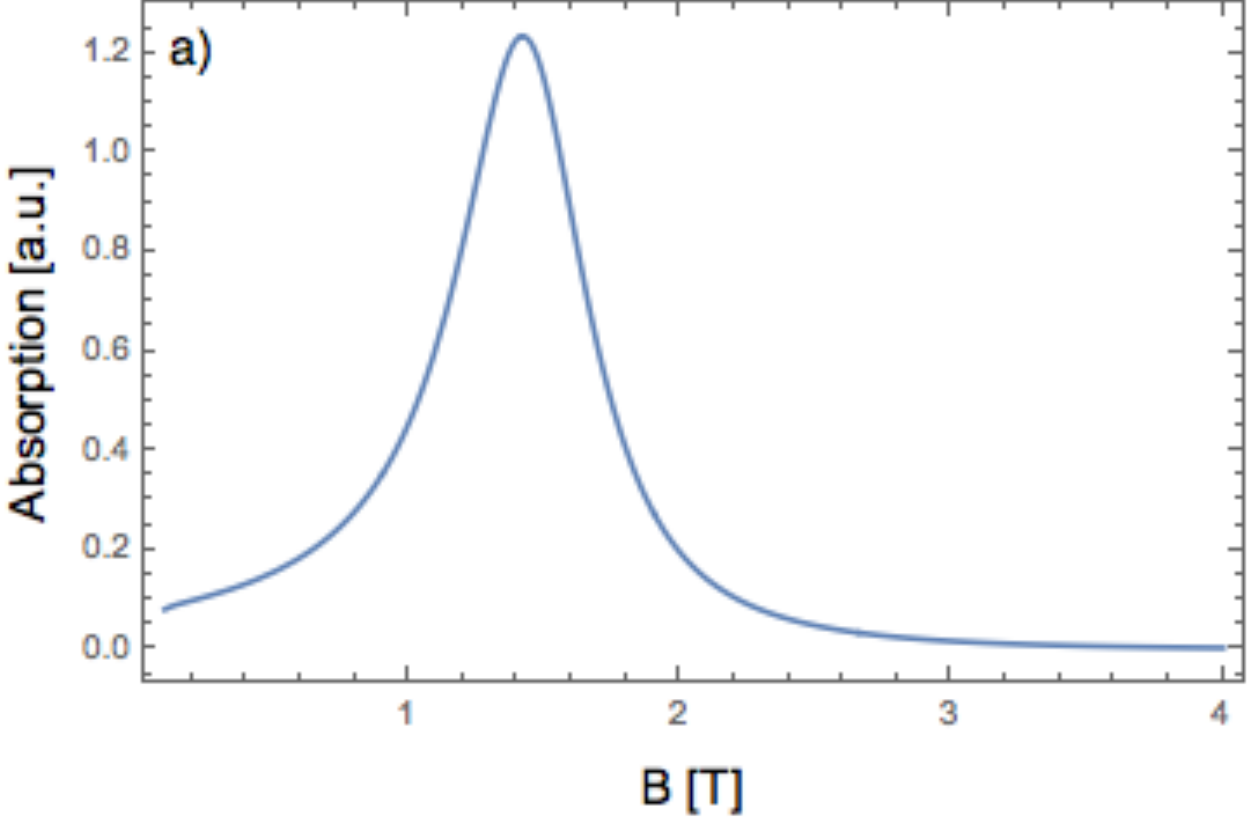}
\includegraphics[width=6cm]{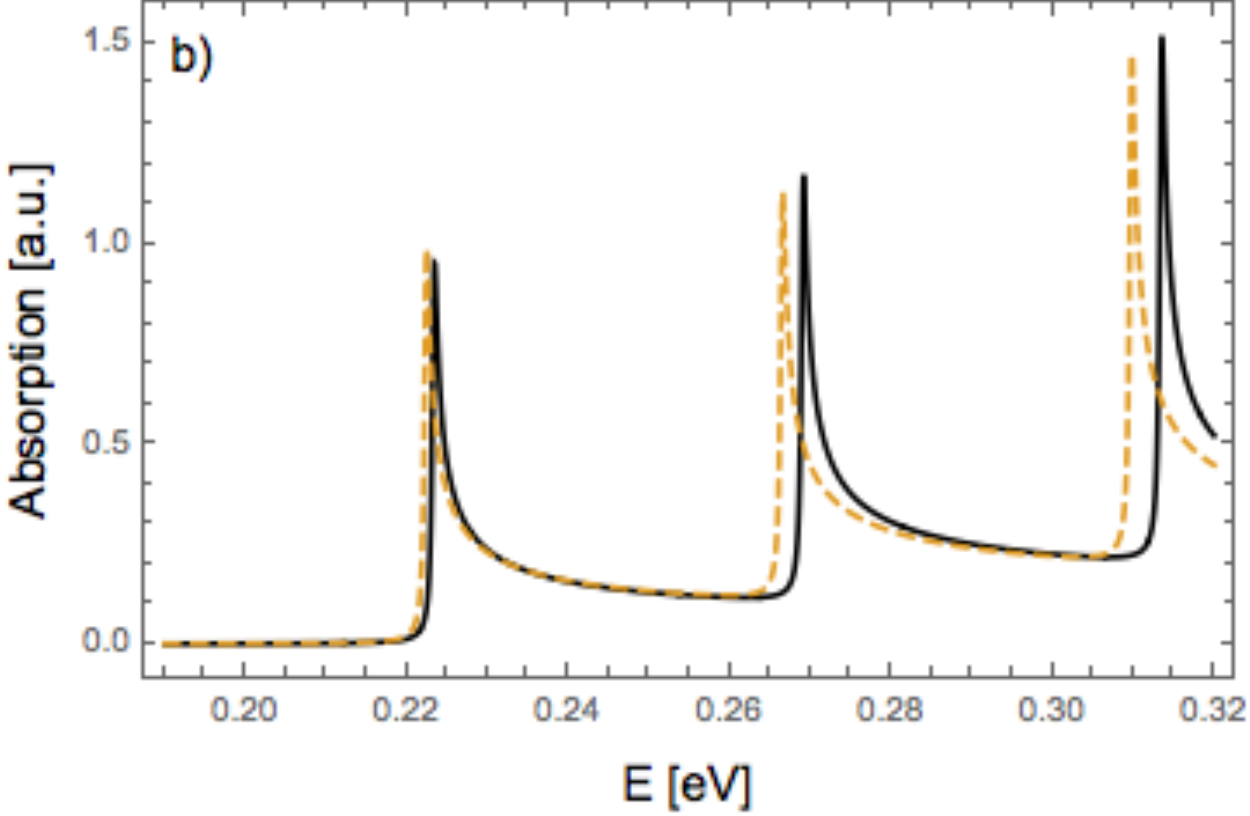}
\caption{\label{fig:electric}
(a)~Cyclotron resonance spectrum of bulk $n$-doped Bi$_2$Se$_3$
as a function of the magnetic field at fixed microwave frequency
$f=\omega/(2\pi)=300\:\mbox{GHz}$.
The electron concentration $n_\rme=10^{17}\:\mbox{cm}^{-3}$,
the temperature is taken 5~K.
(b)~Optical absorption spectrum of bulk undoped Bi$_2$Se$_3$
as a function of the photon energy $E=\hbar\omega$
at fixed $B_z=30$~T. The solid and dashed curves correspond
to the left and right circular polarizations.
The band structure parameters are taken from \cite{Orlita2015}
and \cite{Liu2010} (see main text for details).
}
\end{figure}

In Figures~\ref{fig:TransitionsE},~\ref{fig:electric},
we show allowed transitions for a few Landau levels,
as well as the corresponding absorption spectra,
calculated using~\eref{FGR=} with the energies and
matrix elements obtained by numerically diagonalizing
\eref{Hn=} for each~$k_z$, without employing the parabolic
approximation for $E_{nj}(k_z)$. 
We use the parameters given in \cite{Orlita2015} for
Bi$_2$Se$_3$: $C_2=3\:\mbox{eV}\cdot\mbox{\AA}^2$,
$M_0=0.1\:\mbox{eV}$, $M_2=-22.5\:\mbox{eV}\cdot\mbox{\AA}^2$,
$\mathcal{A}_0=3.1\:\mbox{eV}\cdot\mbox{\AA}$, 
$g_{1z}=g_{2z}=0$, which determine the peak positions.
The peak widths are determined by the $k_z$~dependence of the
energies, and we take the corresponding parameters
from \cite{Liu2010}: $C_1=5.7\:\mbox{eV}\cdot\mbox{\AA}^2$,
$M_1=-6.9\:\mbox{eV}\cdot\mbox{\AA}^2$,
$\mathcal{B}_0=2.3\:\mbox{eV}\cdot\mbox{\AA}$.
To cut off the square-root singularities, we replace the
$\delta$~function in \eref{FGR=} by a Lorentzian with full
width at half maximum $2\gamma$, setting $\gamma=0.5\:\mbox{meV}$.

\subsection{Magnetic dipole perturbation}

The corresponding perturbation contains three terms:
\begin{equation}\label{Fm=}
F_\mathrm{m}=-\frac{eB_x}{2c}\,zv_y+\frac{eB_x}{2c}\,yv_z
+\frac{\mu_\mathrm{B}B_x}{2}\pmatrix{
0 & 0 & g_{1p} & 0 \cr 0 & 0 & 0 & g_{2p} \cr
g_{1p} & 0 & 0 & 0 \cr 0 & g_{2p} & 0 & 0}.
\end{equation}
The matrix elements of the first term (which we denote by
$F_\mathrm{m1}$) are found using the separability of wave
functions~(\ref{wf4=}).
For the transverse motion, the treatment of $v_y$ is totally
analogous to that of $v_x$ for the electric dipole case,
while the matrix element of $z$ is more conveniently
evaluated if one imposes hard wall boundary conditions at
$z=\pm{L_z}/2$ and uses the standing waves
$\sin{k}_zz$, $\cos{k}_zz$
instead of the propagating ones $e^{ik_zz}$. After some
algebra~\cite{LyMaster}, the resulting contribution of
$F_\mathrm{m1}$ to the absorption can be represented as
\begin{equation}\label{Peinplane=}
P_{\mathrm{e}\perp}(\omega)=2\Re\sigma_{yy}(\omega)\,L_xL_y
\int_{-L_z/2}^{L_z/2}|\mathcal{E}_y(z)|^2\rmd{z},
\end{equation}
where the transverse electric field
$\mathcal{E}_y(z)=-(\omega{z}/2c)B_x$ is the one given by the
Faraday's law. The selection rules are the same as for the
electric dipole case, (\ref{electricSR=}).

In the second term of (\ref{Fm=}), $v_z$ contains (i)~terms
originating from $\partial\varepsilon/\partial{k_z}$,
$\partial\mathcal{M}/\partial{k_z}$, and (ii)~a term
$\propto\mathcal{B}_0$. The former ones are proportional to~$k_z$,
so we neglect them as we are interested in transitions at $k_z=0$.
The term~(ii) can be combined with the third term in (\ref{Fm=}),
to give
\begin{equation}
F_\mathrm{m2}=\frac{eB_x}{2m_0\hbar{c}}\pmatrix{
0 & m_0\mathcal{B}_0y & g_{1p}\hbar^2/2 & 0 \cr
m_0\mathcal{B}_0y & 0 & 0 & g_{2p}\hbar^2/2 \cr
g_{1p}\hbar^2/2 & 0 & 0 & -m_0\mathcal{B}_0y \cr 
0 & g_{2p}\hbar^2/2 & -m_0\mathcal{B}_0y & 0}.
\end{equation}
$F_\mathrm{m2}$ induces transitions between sectors with different
$\sigma$, so it does not interfere with the $\sigma$-conserving
$F_\mathrm{m1}$.
When calculating the matrix elements
of $F_\mathrm{m2}$, we encounter
\begin{equation}\label{nyn=}
\langle\phi_n|y|\phi_{n'}\rangle=k_xl_B^2\delta_{nn'}
-l_B\sqrt{\frac{n'}2}\,\delta_{n',n+1}
-l_B\sqrt{\frac{n}2}\,\delta_{n,n'+1}.
\end{equation}
The term $k_xl_B^2\delta_{nn'}$ gives rise to absorption which can
be identified as
\begin{equation}\label{Pez=}
P_{\mathrm{e}z}(\omega)=2\Re\sigma_{zz}(\omega)\,L_xL_z
\int_{-L_y/2}^{L_y/2}|\mathcal{E}_z(y)|^2\rmd{y},
\end{equation}
with $\mathcal{E}_z(y)=(\omega{y}/2c)B_x$ from the Faraday's law
(indeed, $k_xl_B^2$ is the $y$~coordinate of the cyclotron orbit
guiding center). The intraband matrix elements of the $\delta_{nn'}$
term vanish at $k_z=0$ because of the matrix structure; this term
produces a finite intraband absorption
only when combined with the first order in~$k_z$, which flips the
spin back. The resulting intraband absorption is represented by
the Drude peak at $\omega=0$ (as the electron motion in the $z$
direction remains free), which is beyond the scope of our analysis.
In the interband absorption spectrum, the
$\delta_{nn'}$ term does produce additional peaks.

\begin{figure}
\includegraphics[width=10cm]{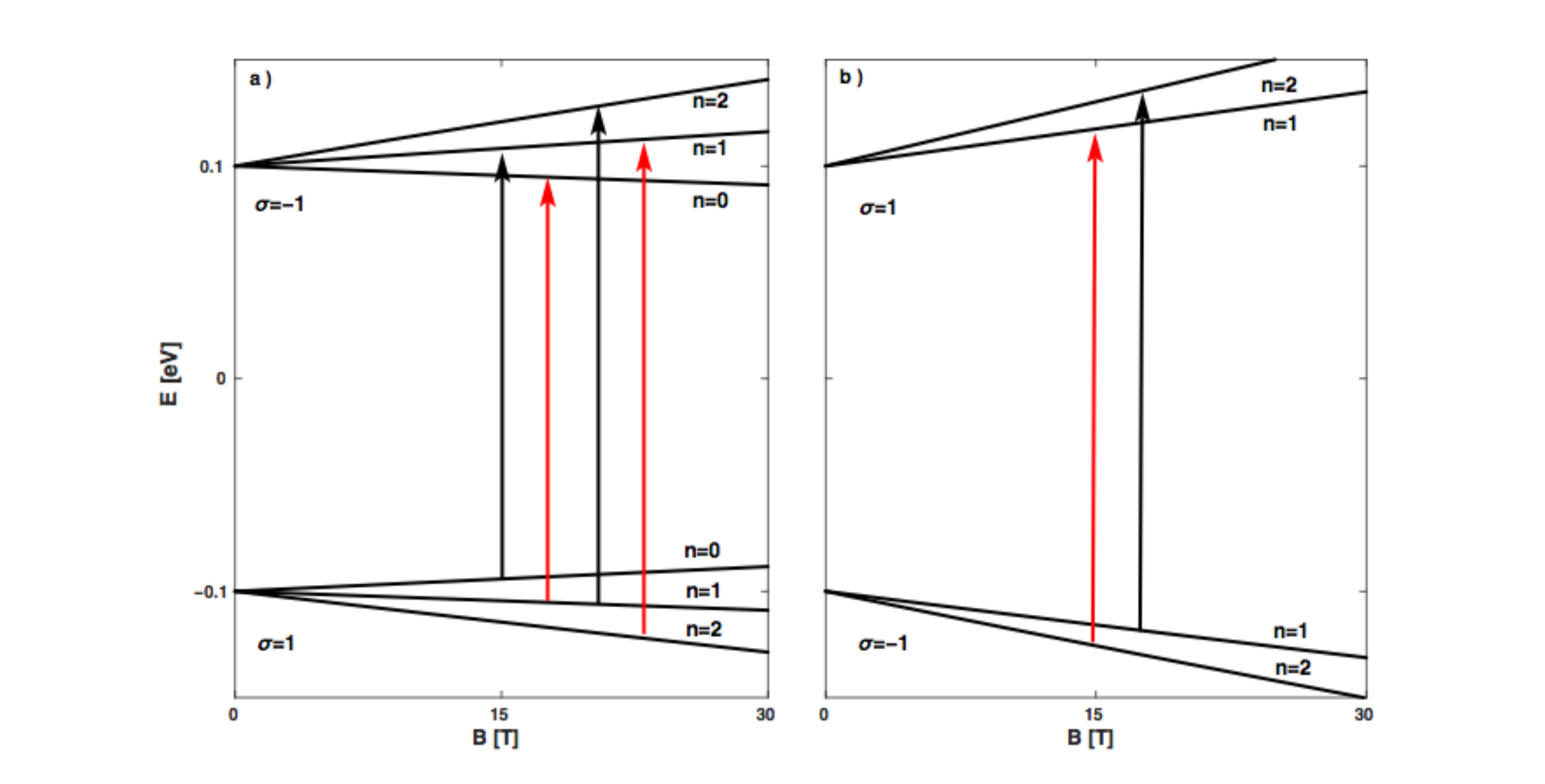}
\caption{\label{fig:TransitionsM}
Same as in Figure~\ref{fig:TransitionsE}, but magnetic dipole
transitions are shown.
}
\end{figure}

\begin{figure}
\includegraphics[width=6cm]{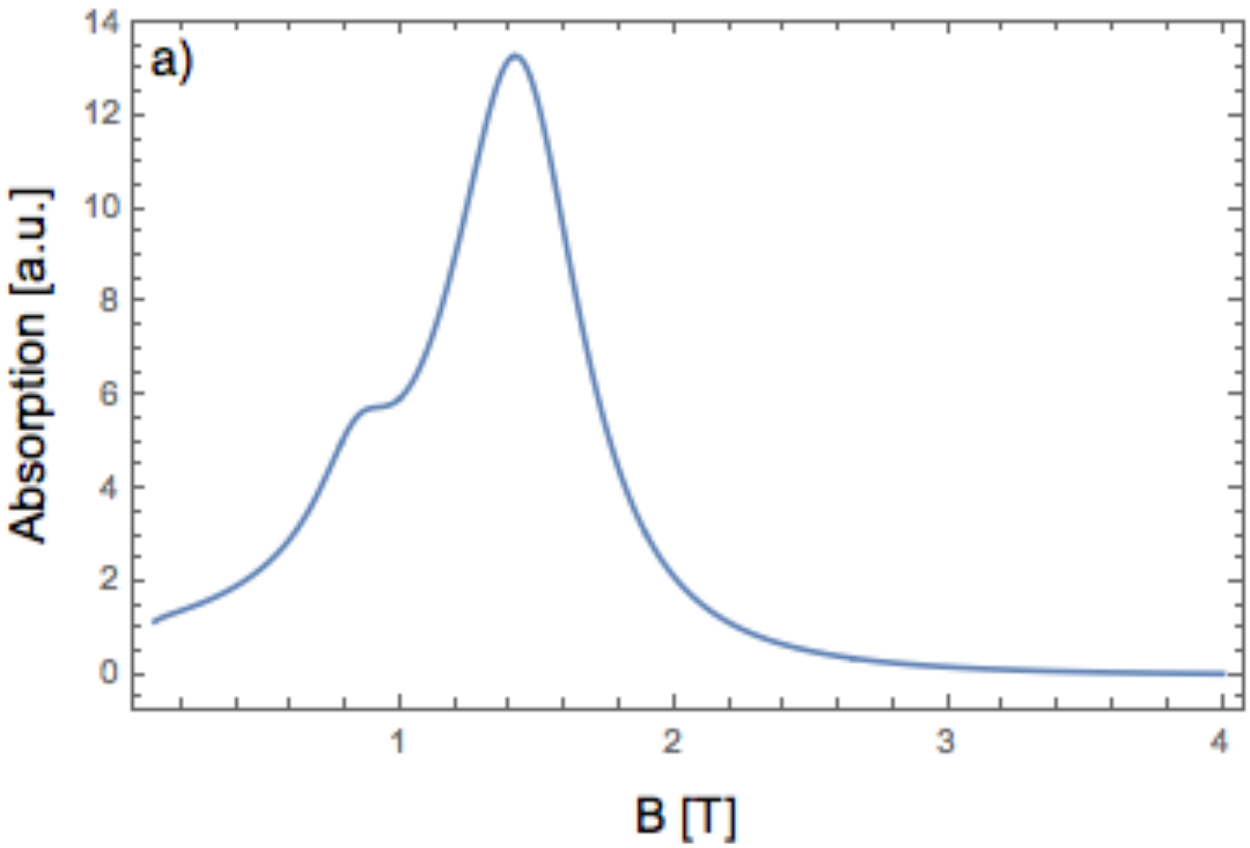}
\includegraphics[width=6cm]{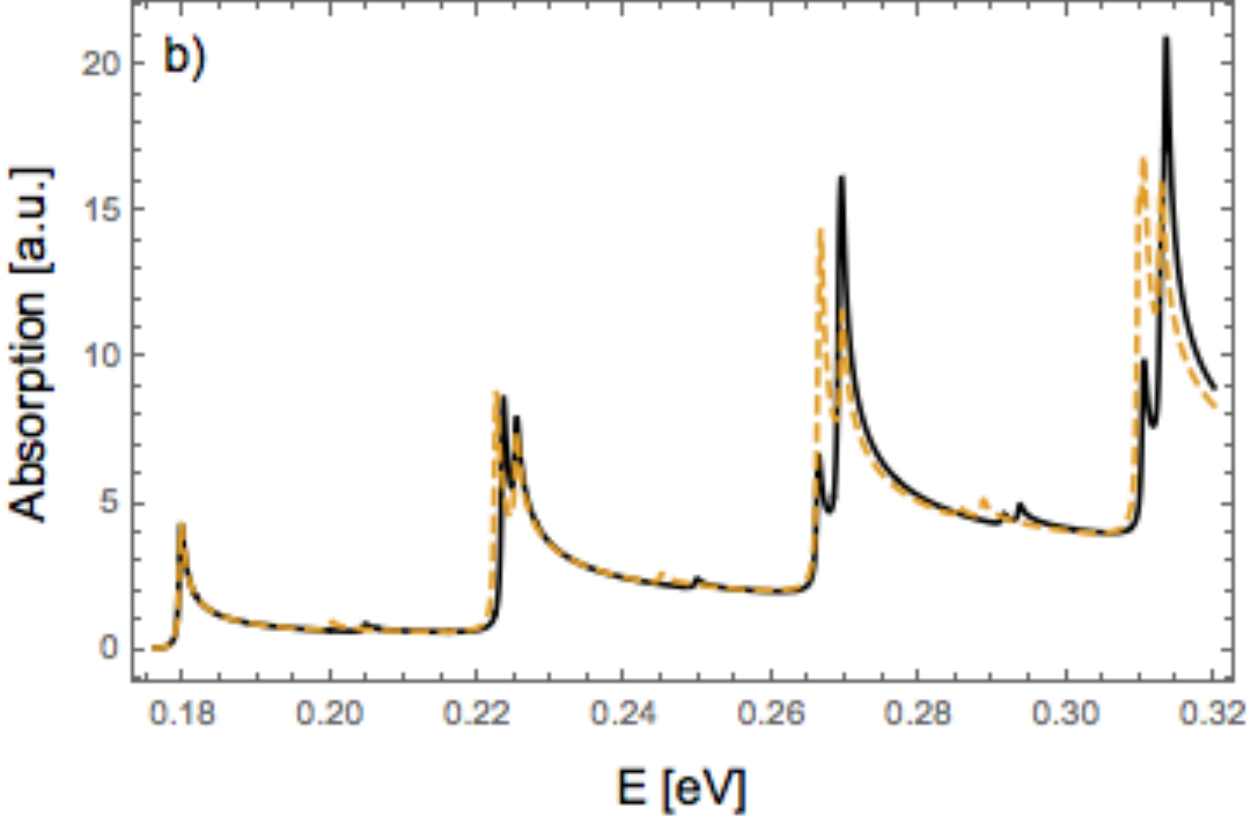}
\caption{\label{fig:magnetic}
(a)~Intraband absorption (ESR) spectrum of an $n$-doped
Bi$_2$Se$_3$ sample of thickness $L_z=500\:\mathrm{nm}$.
(b) Interband absorption spectrum of a Bi$_2$Se$_3$
sample with size $L_{x,y,z}=100\:\mathrm{nm}$
subject to a constant magnetic field $B_z=30\:\mbox{T}$ and
and an oscillating transverse magnetic field.
All other paramaters and notations are the same as for
Figure~\ref{fig:electric}.
}
\end{figure}

Only the last two terms in~(\ref{nyn=}) produce
new transitions which are not contained in $\sigma_{ij}(\omega)$,
and thus can be associated with magnetic dipole. The corresponding 
selection rules are
\begin{equation}\label{magneticSR=}
n,\sigma,\lambda\to n\pm{1},-\sigma,\lambda'.
\end{equation}
%The absorption due to the magnetic dipole transitions can be
%related to the magnetic susceptibility $\mu_{ij}(\omega)$ as%
%\footnote{
%Note, however, that the same absorption can be represented as the
%response to the electric field \emph{gradient} (as the latter is
%rigidly related to the magnetic field by Faraday's law), and thus
%described in terms of the conductivity
%$\sigma_{ij}(\omega,\mathbf{k})$,
%depending on the wave vector~$\mathbf{k}$. This is a manifestation
%of the well-known fact: if spatial dispersion of the dielectric
%susceptibility is included, the magnetic susceptibility can be set
%to $\mu_{ij}(\omega)=\delta_{ij}$ without any loss of generality
%\cite{LL8}.}
%\begin{equation}
%P_\mathrm{m}(\omega)=\frac{1}{2\pi}\Im\mu_{xx}(\omega)
%|B_x|^2L_xL_yL_z.
%\end{equation}
We show these magnetic transitions for a few Landau
levels in figure~\ref{fig:TransitionsM} as well as the absorption 
spectrum in figure~\ref{fig:magnetic}.
The spectrum contains (i)~the same peaks as in
figure~\ref{fig:electric}, due to contribution~(\ref{Peinplane=})
from the transverse electric dipole,
(ii)~peaks due to the $z$~component of the electric dipole,
contribution~(\ref{Pez=}), whose frequencies turn out to be very
close to those of the previous series, and
(iii)~magnetic peaks, seen as very weak features near the energies
0.20, 0.25, 0.29~eV. 
The relative intensity of the weak third series with respect to
the two first ones depends on the sample size.
Indeed, for a given strength of the transverse magnetic
field~$B_x$, the typical values of the electric field components
in the sample are $\mathcal{E}_y~\sim(\omega{L}_z/c){B}_x$,
$\mathcal{E}_z~\sim(\omega{L}_y/c){B}_x$
[as \eref{magdip=} assumes the sample to be placed in the
node of the electric field].
Thus, quite a small value $L_{x,y,z}=100\:\mbox{nm}$ was
chosen to calculate the interband spectra in
Figure~\ref{fig:magnetic}, in order for the magnetic peaks to be
noticeable. Some estimates for the relative intensities are
given in the next section.

\section{Transitions' classification and the $g$ factors}
\label{sec:Discussion}

For the subsequent discussion we have to specify the sign of the gap 
parameter~$M_0$. In fact, this sign is just a matter of convention.
Indeed, if in the Hamiltonian \eref{Hk=}, \eref{HZ=} one changes
simultaneously the signs of $M_{0,1,2}$ and exchanges
the inner and outer blocks (the two sectors with different~$\sigma$)
which also implies the exchange
$g_{1z}\leftrightarrow{g}_{2z}$, $g_{1p}\leftrightarrow{g}_{2p}$,
the Hamiltonan remains intact. We prefer the $M_0>0$ convention, as
then Hamiltonian \eref{Hk=} has the same structure as the Dirac
Hamiltonian in quantum electrodynamics~\cite{LL4}. In fact, the
Dirac Hamiltonian is given by \eref{Hk=} with
$\varepsilon(\mathbf{k})=0$,
$\mathcal{M}(\mathbf{k})=m_0\hbar^2{c}^2$,
$\mathcal{A}_0=\mathcal{B}_0=\hbar{c}$.
In the following, we will identify different transitions using
the analogy with the non-relativistic limit of the Dirac
Hamiltonian, which corresponds to the low-energy or low-$B_z$ limit
$|E_{n,\sigma,\lambda}|-M_0\ll{M}_0$.

If $M_0>0$, then $n=0$, $\sigma=-1$ ($\downarrow$) level
is associated with the conduction band ($\lambda=+1$),
and $n=0$, $\sigma=+1$ ($\uparrow$)
with the valence band ($\lambda=-1$). To match the non-relativistic
limit, we shift the $n$ index down by 1 for half of the levels:
$\tilde{n}=n-\lambda\sigma/2-1/2$,
%\begin{eqnarray}
%\sigma=\downarrow,\;\lambda=+1\;\mbox{or}\;
%\sigma=\uparrow,\;\lambda=-1:&&\;\tilde{n}=n,\\
%\sigma=\uparrow,\;\lambda=+1\;\mbox{or}\;
%\sigma=\downarrow,\;\lambda=-1:&&\;\tilde{n}=n-1,
%\end{eqnarray}
so that for each $\tilde{n}\geqslant{0}$ we have four levels with
$\sigma=\pm{1}$, $\lambda=\pm{1}$. 

First, consider intraband transitions. The electric dipole
selection rule (\ref{electricSR=}) remains the same in the new
representation: $\tilde{n},\sigma\to\tilde{n}\pm{1},\sigma$.
These transitions are associated with the cyclotron resonance.
The magnetic dipole selection rules \eref{magneticSR=} in the
new representation are listed in Table~\ref{tab:msr}. Since
a left/right circularly polarized photon carries an angular
momentum of $\pm{1}$, we can associate $\tilde{n}$ and
$\sigma/2$ with the effective orbital and spin angular momentum
for electrons in the conduction band. For holes in the valence
band, these would be $-\tilde{n}$ and $\sigma/2$, respectively.
The transitions $\tilde{n},\sigma\to\tilde{n},-\sigma$ can be
identified with the electron spin resonance. For the Dirac
Hamiltonian, these transitions correspond to the electron spin
flip in the non-relativistic limit, while the transitions with
$\tilde{n}$ changing by $\pm{2}$ are due to spin-orbit coupling,
and thus are weak. 
Thus, it is natural to define the effective
$g$-factors for electrons in the conduction band and for the
holes in the valence band as
\numparts\begin{eqnarray}
g_e=\frac{1}{\mu_\mathrm{B}}\left.
\frac{\partial(E_{n,\downarrow,+1}-E_{n+1,\uparrow,+1})}%
{\partial{B}_z}\right|_{B_z=0}=
-\frac{2m_0\mathcal{A}_0^2}{M_0}-g_{1z},\\
g_h=\frac{1}{\mu_\mathrm{B}}\left.
\frac{\partial(-E_{n+1,\downarrow,-1}+E_{n,\uparrow,-1})}%
{\partial{B}_z}\right|_{B_z=0}
=\frac{2m_0\mathcal{A}_0^2}{M_0}-g_{2z}.
\end{eqnarray}\endnumparts
Note that $g_e<0$ because the electron charge is negative and
$\mu_\mathrm{B}>0$, while $g_h>0$ as the hole has a positive
charge. The numerical values from~\cite{Orlita2015} give
$g_h=-g_e\approx{2}6$.

\begin{table}
\begin{tabular}{|c|c|c|}
\hline & $\circlearrowleft$ & $\circlearrowright$ \\
\hline \parbox{1.7cm}{\vspace*{2mm} conduction band\vspace*{2mm}} &
\parbox{2.2cm}{$\tilde{n},\downarrow\;\to\;\tilde{n},\uparrow$\\
$\tilde{n},\uparrow\;\to\;\tilde{n}+2,\downarrow$}
& \parbox{2.2cm}{$\tilde{n},\uparrow\;\to\;\tilde{n},\downarrow$\\
$\tilde{n},\downarrow\;\to\;\tilde{n}-2,\uparrow$}\\
\hline \parbox{1.7cm}{\vspace*{2mm} valence band \vspace*{2mm}} &
\parbox{2.2cm}{$\tilde{n},\uparrow\;\to\;\tilde{n},\downarrow$\\
$\tilde{n},\downarrow\;\to\;\tilde{n}+2,\uparrow$}
& \parbox{2.2cm}{$\tilde{n},\downarrow\;\to\;\tilde{n},\uparrow$\\
$\tilde{n},\uparrow\;\to\;\tilde{n}-2,\downarrow$}\\ \hline
\end{tabular}
\caption{Magnetic dipole selection rules for intraband transitions.
The left and right circular polarizations are labeled by
$\circlearrowleft$ and $\circlearrowright$, respectively.}
\label{tab:msr}
\end{table}

As intraband transitions are usually probed at low magnetic fields,
the relative strengths of different transitions can be reliably
estimated in the ``non-relativistic limit'' using the expansion
in small parameters
$\sqrt{n}\mathcal{A}_0/(l_BM_0)\ll{1}$, $nM_2/(l_B^2M_0)\ll{1}$.
By evaluating the ratio
of the squares of the corresponding matrix elements at $k_z=0$,
we find that the ESR transitions are weaker than the cyclotron
resonance by a factor
$\sim(1/8n)(B_x/\mathcal{E}_x)^2(\mathcal{B}_0/\mathcal{A}_0)^2
(\omega{l}_B/c)^2$. The transitions with $\Delta\tilde{n}=\pm{2}$
are weaker than the ESR ones by an additional factor
\begin{equation}\label{eta=}
\eta\sim\left(n\,\frac{\mathcal{A}_0^2+4M_0M_2}{2M_0^2l_B^2}\right)^2
\end{equation}
(note the strong cancellation between the two terms in the
numerator for the parameters of Bi$_2$Se$_3$ \cite{Orlita2015}).

For transitions from the valence to the conduction band, the
dominant transverse electric dipole transitions in the
non-relativistic limit are those with $\tilde{n}\to\tilde{n}$,
occurring in $\sigma=+1$ and $\sigma=-1$ sectors for the left
and right circular polarizations, respectively. They are
accompanied by transitions with $\Delta\tilde{n}=\pm{2}$, whose
intensity is weaker at low fields by the same factor~$\eta$ as
in \eref{eta=} [note though that the relevant magnetic
fields are much higher than for the intraband transitions, and
they are out of the ``non-relativistic'' limit, so
\eref{eta=} provides only an order-of-magnitude estimate].
Among the magnetic dipole interband transitions,
$\tilde{n},\sigma\to\tilde{n}\pm{1},-\sigma$, there is  no
dominant series, they all have similar strength. The ratio between
the strength of the interband magnetic dipole transitions and the
leading series of the transverse electric dipole transitons can be
estimated
as $\sim(n/8)(B_x/\mathcal{E}_x)^2(\mathcal{B}_0/\mathcal{A}_0)^2
(\omega{l}_B/c)^2$, also similar to that in the intraband case.
As both are proportional to $l_B^2\propto{1}/B_z$, the magnetic
dipole peaks are more easily observed on top of the electric
dipole ones for the intraband transitions than for the interband
ones. Hence, a larger size is chosen for the former ones than
for the latter to calculate the spectra shown in
Figure~\ref{fig:magnetic}.
The ratio between the transverse and $z$ electric dipole transitons
is simply
$(\mathcal{E}_x/\mathcal{E}_z)^2(\mathcal{A}_0/\mathcal{B}_0)^2$.

\section{Conclusions}
\paragraph{}

We have studied the bulk response of topological insulators
Bi$_2$Se$_3$, Bi$_2$Te$_3$ and Sb$_2$Te$_3$ in a static magnetic 
field to electric dipole and magnetic dipole perturbations. It
corresponds to cyclotron resonance, electron spin resonance and
interband optical absorption. Using the effective four-band model
\cite{Zhang2009,Liu2010}, we found the Landau levels in the static
magnetic field, and calculated the energy absorption spectrum
when an oscillating electric or magnetic field is applied
perpendicular to the static magnetic field. We have determined the
corresponding selection rules; for intraband transitions they are
the same as for genuine
Dirac electrons in quantum electrodynamics. From these selection
rules, we were able to separate the effective orbital and spin 
degrees of freedom and deduce the effective $g$~factors for
electrons and holes.

\ack
The authors are grateful to M. Orlita and M. Potemski for very 
stimulating discussions and critical reading of the manuscript.
O.~Ly thanks Laboratoire de Physique et Mod\'elisation des 
Milieux Condens\'es for hospitality and support during his 
undergraduate internship when this work started.

\end{document}